# Kochen-Specker for many qubits and the classical limit.

Alejandro A. Hnilo

*CEILAP, Centro de Investigaciones en Láseres y Aplicaciones, (CITEDEF-CONICET);*
*J.B. de La Salle 4397, (1603) Villa Martelli, Argentina.*
*email: alex.hnilo@gmail.com*

*Abstract.*

Several arguments demonstrate the incompatibility between Quantum Mechanics and classical Physics. Bell's inequalities and Greenberger-Horne-Zeilinger (GHZ) arguments apply to specific non-classical states. The Kochen-Specker (KS) one, instead, is especially appealing for it applies to any state. Nevertheless, in spite of the incompatibility, quantum predictions must converge to classical ones as the macroscopic scale is approached. This convergence is known as "classical limit", and is difficult to explain within quantum formalism. In this short paper, the simplified Mermin-Peres form (two qubits) of the KS argument is extended to an arbitrary number of qubits. It is shown that quantum and classical predictions converge as the number of qubits is increases to the macroscopic scale. This way to explain the classical limit concurs with, and improves, a result previously reported for GHZ states. The demonstration for the general case (i.e., for all possible observables) that the classical limit is the consequence of merely increasing the number of particles, is important and seems to be at hand.

*November 25th, 2024.*

The incompatibility between Quantum Mechanics (QM) and classical Physics is evident in the difference of their predictions for some selected situations. Bell's inequalities (BI) [1] and Greenberger-Horne-Zeilinger (GHZ) [2] arguments demonstrate the incompatibility for specific quantum states. The Kochen-Specker one (KS) instead, [3] is especially valuable for it applies to *any* state.

In spite of that essential incompatibility, QM predictions must converge to the ones of classical mechanics as the observed system approaches the macroscopic scale. This convergence is known as *classical limit*. It is often expressed in terms of a vanishing Planck's constant ($h \to 0$) as scaled with the system's action. This makes operators to commute and Hamilton's principle to take its classical expression. If movement of particles is involved, classical trajectories are retrieved. Typical QM effects as superposition, interference and entanglement disappear. However, the classical limit is not easy to explain within QM. The "canonical" explanation assumes the macroscopic system to unavoidably interact with the environment, which is assumed Markovian. Quantum coherence is hence shared with an increasingly larger number of degrees of freedom and, eventually, lost. Alternative explanations have been proposed, f.ex.: rigged Hilbert spaces as the framework of QM [4], decay of coherence induced by an interaction of the massive system with the gravity field [5,6], decay of coherence because of hypothesized new physics that scales with the system's mass [7] and the idea that no decay occurs at all and parallel realities exist [8]. A review of alternative explanations can be found in [9].

A general, intuitive definition of the classical limit can be stated as follows: *the classical limit is reached when all observations on the system can be explained by a classical model*. This definition is independent of the system's action and the existence of trajectories. A similar strategy based on expectation values of operators rather than trajectories (what is especially relevant in quantum chaos), is adopted in [10,11]. In [12] it is demonstrated this form of the classical limit to be reached by merely increasing the number $q$ of qubits in a GHZ state, with no need to suppose an environment, or a gravity field, or new physics, etc. Yet, this demonstration has two arguable points: *i)* GHZ states are assumed to be the most non-classical states possible (what is often believed, but remains to be demonstrated); *ii)* some (small) imperfection $\varepsilon > 0$ in the instruments is assumed to exist. Under these assumptions the difference between QM and classical predictions vanishes as $(1-2\varepsilon)^q \to 0$, implying that the classical limit, as it is defined above, is reached by merely assuming $q \gg 1$. Take into account that, at the macroscopic scale, the value of $q$ can be of the order of Avogadro's number.

In this paper, it is demonstrated that the KS argument supports a similar explanation of the classical limit. It improves the result obtained for GHZ states, for the KS argument is independent of the state (what removes the first arguable point) and no instrumental imperfection is assumed (what removes the second arguable point). On the other hand, the demonstration is valid for a usual and natural, but specific, observable.

The original KS argument involves a spin-1 system and 117 different orientations in space. After intricate geometrical reasoning, it is demonstrated that there is no way to assign values to all these observables before measurement in a way consistent with QM predictions. Mermin [13] and Peres [14] showed the same conclusion can be reached in a simpler way by going up one dimension, to a system of two qubits. I review their argument. Consider the array [15]:

| $\sigma_z^{(1)}$ | $\sigma_z^{(2)}$ | $\sigma_z^{(1)}.\sigma_z^{(2)}$ |
|---|---|---|
| $\sigma_x^{(2)}$ | $\sigma_x^{(1)}$ | $\sigma_x^{(1)}.\sigma_x^{(2)}$ |
| $\sigma_z^{(1)}.\sigma_x^{(2)}$ | $\sigma_x^{(1)}.\sigma_z^{(2)}$ | $\sigma_y^{(1)}.\sigma_y^{(2)}$ |

(1)

where $\sigma_j^{(i)}$ is the j-Pauli matrix acting on qubit *i*. In each of the rows or columns the observables are mutually commuting, so that they can be measured in any order without perturbing the others or the final result. At first sight, this situation allows a classical description. Let suppose then that it is possible to assign an outcome value (+1 or -1) to the elements in the array before measurement. That this assignment (*any* assignment) cannot be consistent with QM predictions is easy to demonstrate. Let multiply all columns $C_k$ and rows $R_k$ in the array, then one expects (the identity matrix in the rhs is omitted in what follows):

$$\Pi_k C_k \times R_k = 1 \quad (2)$$

because each element in the array appears twice in the product. According to QM instead, the result for all rows and columns is +1 excepting for the third column:

$$\sigma_z^{(1)}.\sigma_z^{(2)}.\sigma_x^{(1)}.\sigma_x^{(2)}.\sigma_y^{(1)}.\sigma_y^{(2)} = \sigma_z^{(1)}.\sigma_x^{(1)}.\sigma_y^{(1)}.\sigma_z^{(2)}.\sigma_x^{(2)}.\sigma_y^{(2)} = -1 \quad (3)$$

so that:

$$\Pi_k C_k \times R_k = -1 \quad (4)$$

Therefore, the supposed assignment of outcomes (no matter which one) previous to actual measurement cannot fit the QM predictions. Note the state of two qubits (on which the Pauli matrices would operate) has not been defined. The result is valid for *any* state. This is one of the strongest arguments against classical counterfactual definiteness [1].

Let briefly see how it works. The outcome +1 in the two first columns and rows is evident. The third row reads:

$$\sigma_z^{(1)}.\sigma_x^{(2)}.\sigma_x^{(1)}.\sigma_z^{(2)}.\sigma_y^{(1)}.\sigma_y^{(2)} = \sigma_z^{(1)}.\sigma_x^{(1)}.\sigma_y^{(1)}.\pmb{\sigma_x^{(2)}.\sigma_z^{(2)}}.\sigma_y^{(2)} = +1 \quad (5)$$

The difference with the third column is because here is a $\sigma_x.\sigma_z = -i\sigma_y$ factor (in red and bold type) which is missing in eq.3. One may wonder if this can be changed, for all the elements in eq.1 commute with each other. Examination of the arrays' structure shows that this is impossible. In the third column, the $\sigma_z^{(j)}$ (j=1,2) are *both* before or after the corresponding $\sigma_x^{(j)}$ (j=1,2), what always produces a factor -1. In the third row instead, if $\sigma_z^{(1)}$ is chosen to be *before* $\sigma_x^{(1)}$, then $\sigma_z^{(2)}$ will appear *after* $\sigma_x^{(2)}$, or vice versa. This produces an additional minus sign (total factor $-i^2$) and hence $R_3 = +1$ and $\Pi_k C_k \times R_k = -1$.

The KS argument demonstrates logical contradiction but, in order to test it experimentally, an inequality involving average results of observed outcomes is necessary. The following bound has been demonstrated for the array in eq.1 [16]:

$$X_{KS} = \sum_{j=1}^{3} \langle R_j \rangle + \langle C_1 \rangle + \langle C_2 \rangle - \langle C_3 \rangle \leq 4 \qquad (6)$$

while QM predicts $X_{KS} = 6$ (the total number of rows and columns). The value $X_{KS} \approx 5.46 > 4$ has been experimentally observed [15].

A generalized (quite involved) KS argument for a broad class of hidden variable theories has been recently published [17]. The extension of the Mermin-Peres argument to an arbitrary number $q$ of qubits, instead, is simple. The array that extends eq.1 has three rows and $q+1$ columns. As in eq.1, the elements in the third row (excepting the last element) are the product of the elements in the same column, and the elements in the last column (excepting the last element) are the products of the elements in the same row; the remaining element is the product of the $\sigma_y^{(k)}$ of all qubits:

| $\sigma_z^{(1)}$ | ... | $\sigma_z^{(k)}$ | ... | $\sigma_z^{(q)}$ | $\sigma_z^{(1)}...\sigma_z^{(q)}$ |
|---|---|---|---|---|---|
| $\sigma_x^{(k'\neq 1)}$ | ... | $\sigma_x^{(k'\neq k)}$ | ... | $\sigma_x^{(k'\neq q)}$ | $\sigma_x^{(1)}...\sigma_x^{(q)}$ |
| $\sigma_z^{(1)}.\sigma_x^{(k'\neq 1)}$ | ... | $\sigma_z^{(k)}.\sigma_x^{(k'\neq k)}$ | ... | $\sigma_x^{(k'\neq q)}.\sigma_z^{(q)}$ | $\sigma_y^{(1)}...\sigma_y^{(q)}$ |

$$(7)$$

Note that the order of the $\sigma_x^{(k')}$ in the second row is different from the order of the $\sigma_z^{(k)}$ in the first row, so that in no column (excepting $C_{q+1}$) $\sigma_z$ and $\sigma_x$ operate on the same qubit. This makes the commutations as in eq.1 to remain valid. In this array $R_1=R_2=1$ too, for they are the product ($q$ times) of squares of the corresponding Pauli matrices ($\sigma_z$ or $\sigma_x$). The columns are also the product (twice) of squares of Pauli matrices and are hence equal to +1, excepting the last column $C_{q+1} = (\sigma_z.\sigma_x.\sigma_y)^q = (i)^q$. It only remains $R_3 = (\sigma_z.\sigma_x.\sigma_y)^m \times (\sigma_x.\sigma_z.\sigma_y)^{q-m} = (i)^m \times (-i)^{q-m}$ where $m$ depends of the order chosen for the $\sigma_x^{(k')}$ in $R_2$ (see below). Therefore:

$$\Pi_k C_k \times R_k = C_{q+1} \times R_3 = (i)^q \times (i)^m \times (-i)^{q-m} = (-1)^{q-m} \times (i)^{2q} \qquad (8)$$

where $m$ is the number of factors $\sigma_z^{(k)}.\sigma_x^{(k)}.\sigma_y^{(k)}$ that appear in $R_3$. In consequence, for any value of $q$, it is always possible to choose an ordering of the $\sigma_x^{(k')}$ in $R_2$ (what defines the value of $m$) so that the rhs in eq.8 is -1, and hence incompatible with any classical assignment of outcomes. Eq.8 extends the results of the KS argument to all values of $q$.

As an example, consider the case $q=4$. A possible array is:

| $\sigma_z^{(1)}$ | $\sigma_z^{(2)}$ | $\sigma_z^{(3)}$ | $\sigma_z^{(4)}$ | $\sigma_z^{(1)}\ldots\sigma_z^{(4)}$ |
|---|---|---|---|---|
| $\sigma_x^{(4)}$ | $\sigma_x^{(3)}$ | $\sigma_x^{(2)}$ | $\sigma_x^{(1)}$ | $\sigma_x^{(1)}\ldots\sigma_x^{(4)}$ |
| $\sigma_z^{(1)}.\sigma_x^{(4)}$ | $\sigma_z^{(2)}.\sigma_x^{(3)}$ | $\sigma_z^{(3)}.\sigma_x^{(2)}$ | $\sigma_z^{(4)}.\sigma_x^{(1)}$ | $\sigma_y^{(1)}\ldots\sigma_y^{(4)}$ |

(9)

in this array, $C_5 = (\sigma_z.\sigma_x.\sigma_y)^4 = (i)^4 = +1$ and $R_3$, after reordering the operators acting on the same qubit: $\sigma_z^{(1)}.\sigma_x^{(1)}.\sigma_y^{(1)}.\sigma_z^{(2)}.\sigma_x^{(2)}.\sigma_y^{(2)}.\sigma_x^{(3)}.\sigma_z^{(3)}.\sigma_y^{(3)}.\sigma_x^{(4)}.\sigma_z^{(4)}.\sigma_y^{(4)} = (i)^2 \times (-i)^2 = +1$ (note that $m=2$). Therefore, the array in eq.9 holds to eq.2 and is not in contradiction with classical physics. In order to retrieve the contradiction, it suffices to modify $R_2$ by swapping the positions of $\sigma_x^{(4)}$ and $\sigma_x^{(2)}$ (in $C_1$ and $C_3$) then $R_3 = \sigma_z^{(1)}.\sigma_x^{(1)}.\sigma_y^{(1)}.\sigma_x^{(2)}.\sigma_z^{(2)}.\sigma_y^{(2)}.\sigma_x^{(3)}.\sigma_z^{(3)}.\sigma_y^{(3)}.\sigma_x^{(4)}.\sigma_z^{(4)}.\sigma_y^{(4)} = (i) \times (-i)^3 = -1$ (note that $m=1$ now). Of course, there is more than one way to get this result.

Depending of the value of $q$, $C_{q+1}$ and $R_3$ can be imaginary. Yet, their product is always real and, as $C_{q+1}$ and $R_3$ commute, $\langle R_3 \rangle . \langle C_{q+1} \rangle = \langle R_3.C_{q+1} \rangle$. An appropriate observable $X_{KS(q)}$ that extends eq.6 is then:

$$X_{KS(q)} = 1 + \langle R_1 \rangle + \langle R_2 \rangle + \sum_{j=1}^{q}\langle C_j \rangle - \langle R_3.C_{q+1} \rangle \le q+2 \qquad (10)$$

while the QM prediction is (for an appropriate choosing of $m$) the total number of rows and columns: $X^{QM}_{KS(q)} = q+4$. An array of classical instructions that saturates the inequality in eq.10 is easy to find, f.ex., one that assigns +1 to all the elements in eq.7. It is now visible that the difference between the classical and QM predictions for the observable $X_{KS(q)}$ vanishes as the number of qubits increases:

$$X_{KS(q)} / X^{QM}_{KS(q)} = (q+2)/(q+4) \to 1 - \frac{2}{q} \to 1 \text{ if } q \gg 1 \qquad (11)$$

Thus the classical limit, as it is defined before, is reached by merely increasing the number of qubits in the system. This result improves the one obtained for GHZ states, for it is valid for any state and does not require assuming any instrumental imperfection. It concurs with the view that systems with infinite degrees of freedom are always classical [18]. On the other hand, the difference between QM and classical predictions decays slower (for $q \gg 1/\varepsilon$), as $q^{-1}$ instead of $(1-2\varepsilon)^q$.

The result above is an important step to explain the classical limit in simple terms but is not fully satisfactory, for it is obtained for the particular observable $X_{KS(q)}$. This observable is built up as an extension of the one defined in [16] to test the KS argument. It is usual and natural, but is not demonstrated to be the one producing the largest difference between QM and classical predictions (not even for $q=2$). In principle at least, other observables may have a less favorable scaling with $q$.

Recall that the stated definition of the classical limit implies that *all* observations on the system must be explained by a classical model. Anyway, summing up the results obtained for GHZ states and here, there is a basis wide enough to hypothesize that the classical limit is indeed reached by merely increasing the number of particles in the system, with no need of any additional assumption. The search for a fully general demonstration (i.e., for all observables of a system of *q* qubits), starting from the result obtained here, appears to be a promising line for future research. An apparent approach is to demonstrate that $X_{KS}$ (eq.6) is the observable that has the largest difference with classical predictions for $q=2$, and then to extend the result by induction. That demonstration would solve the problem of the classical limit in the simplest way.

**Acknowledgements.**

This material is based upon research supported by grants PIP 00484-22 and PUE 229-2018-0100018CO, both from CONICET (Argentina).

**References.**